\documentclass[preprint,showpacs,preprintnumbers,amsmath,amssymb]{revtex4}
\usepackage{afterpage}
\usepackage[dvipdfmx]{graphicx}
\usepackage{color}

\begin{document}

\title{
Deflection angle of light for an observer and source 
at finite distance from a rotating global monopole} 
\author{Toshiaki Ono}
\author{Asahi Ishihara}
\author{Hideki Asada} 
\affiliation{
Graduate School of Science and Technology, Hirosaki University,
Aomori 036-8561, Japan} 
\date{\today}

\begin{abstract} 
By using a method improved with a generalized optical metric, 
the deflection of light 
for an observer and source at finite distance from a lens object 
in a stationary, axisymmetric and asymptotically flat spacetime 
has been recently discussed 
[Ono, Ishihara, Asada, Phys. Rev. D {\bf 96}, 104037 (2017)]. 
In this paper, 
we study a possible extension of this method to an asymptotically nonflat 
spacetime. We discuss a rotating global monopole.
Our result of the deflection angle of light 
is compared with a recent work 
on the same spacetime but limited within the asymptotic source and observer 
[Jusufi et al., Phys. Rev. D {\bf 95}, 104012 (2017)], in which they employ 
another approach proposed by Werner 
with using the Nazim's osculating Riemannian construction method 
via the Randers-Finsler metric. 
We show that the two different methods give the same result 
in the asymptotically far limit. 
We obtain also the corrections to the deflection angle 
due to the finite distance from the rotating global monopole. 
Near-future observations of Sgr A${}^{*}$ will be able to put a bound 
on the global monopole parameter $\beta$ as $1- \beta < 10^{-3}$ 
for a rotating global monopole model, 
which is interpreted as the bound on the deficit angle 
$\delta < 8\times 10^{-4}$ [rad]. 
\end{abstract}


\maketitle

\section{Introduction}
Since the experimental confirmation of the theory of general relativity 
\cite{Intro1} 
succeeded in 1919 \cite{Intro2}, a lot of calculations of 
the gravitational bending of light have been done not only 
for black holes \cite{Intro3} 
but also for other objects such as wormholes and gravitational monopoles 
\cite{Intro4}.
Gibbons and Werner (2008) proposed an alternative way of deriving 
the deflection angle of light \cite{Intro5}. They assumed that the source and 
receiver are located at an asymptotic Minkowskian region and they used 
the Gauss-Bonnet theorem to a spatial domain described by the optical metric, 
for which a light ray is described as a spatial curve. 
Ishihara et al. have recently extended Gibbons and Werner's idea 
in order to investigate finite-distance corrections in 
the small deflection case (corresponding to a large impact parameter case) 
\cite{Intro6} 
and also in the strong deflection limit for which 
the photon orbits may have the winding number larger than unity \cite{Intro7}. 
In particular, the asymptotic receiver and source have not been assumed.
Our method and Werner's one are limited within asymptotically flat spacetimes.

In this paper, 
we discuss an extension of our method 
applied to a rotating global monopole. 
Due to the existence of a deficit solid angle, 
the spacetime is not asymptotically flat. 
A static solution of a global monopole was found in a paper 
by Barriola and Vilenkin \cite{Intro8}. 
According to their model, global monopoles are configurations 
whose energy density decreases with the distance as $r^{-2}$ and 
whose spacetimes exhibit a solid deficit angle given 
by $ \delta = 8 \pi^2 \eta^2$, where $\eta$ is the scale of gauge-symmetry 
breaking. 
Recently, global monopoles have been discussed as spacetimes 
with a cosmological constant, e.g. in \cite{Intro9}. 
Static spherically symmetric composite global-local monopoles have also been 
studied \cite{Intro10}.
Gravitational lensing in spacetimes with a non-rotating global monopole 
has been intensively investigated, for instance by Cheng and Man \cite{Intro11} 
who studied strong gravitational lensing of a Schwarzschild black hole 
with a solid deficit angle owing to a global monopole. 
More recently, it has also been proposed 
that gravitational microlensing by global monopole may even be used to test 
Verlinde's emergent gravity theory \cite{Intro12}.
As mentioned above, 
we investigate a possible extension of our method to 
stationary, axisymmetric spacetimes with a solid deficit angle,
especially in order to examine finite-distance corrections to 
the deflection angle of light. 
The geometrical setups in the present paper are not those 
in the optical geometry, in the sense that the photon orbit has 
a non-vanishing geodesic curvature, though the light ray in 
the four-dimensional spacetime obeys a null geodesic.

This paper is organized as follows. 
Section II discusses a generalized optical metric for a 
rotating global monopole. 
Section III discusses how to define the deflection angle of light 
in a stationary, axisymmetric spacetime with the deficit angle.
In particular, it is shown that the proposed definition of 
the deflection angle is also coordinate-invariant
by using the Gauss-Bonnet theorem.
We discuss also how to compute the gravitational deflection angle of light 
by the proposed method. 
Section IV is devoted to the conclusion. 
Throughout this paper, we use the unit of $G = c = 1$, 
and the observer may be called the receiver in order to avoid a confusion 
between $r_O$ and $r_0$ by using $r_R$.

\section{Generalized optical metric for rotating global monopole}
\subsection{Rotating global monopole} 
By applying the method of complex coordinate transformation, 
an extension of the static global monopole solution to 
a rotating global monopole spacetime was described 
by R. M. Teixeira Filho and V. B. Bezerra in Ref. \cite{RGMP}. 

Its spacetime metric reads 
\begin{align}
ds^2=& g_{\mu\nu}dx^{\mu}dx^{\nu} \notag\\
=&-\Big(1-\frac{2Mr}{r^2+a^2\cos^2\theta}\Big)dt^2 \notag\\
&+\Bigg[\frac{r^2-a^2\Big\{(1-\beta^2)\sin^2\theta-\cos^2\theta\Big\}}
{r^2-2Mr+a^2}
-(1-\beta^2)
\frac{a^2\sin^2\theta\{2Mr-a^2(1-\sin^4\theta)\}}{(r^2-2Mr+a^2)^2}\Bigg]dr^2
\notag\\
&+\beta^2(r^2+a^2\cos^2\theta)d\theta^2 \notag\\
&+\sin^2\theta\frac{\Big[\beta^2r^4+\{1-(1-2\beta^2)\cos^2\theta\}a^2r^2
+2Ma^2r\sin^2\theta+a^4\cos^2\theta(\beta^2\cos^2\theta+\sin^2\theta)\Big]}
{r^2+a^2\cos^2\theta} d\phi^2 \notag\\
&-\frac{4aMr\sin^2\theta}{r^2+a^2\cos^2\theta}dtd\phi
+2(1-\beta^2)\frac{a\Big\{r^2\sin^2\theta-a^2\cos^2\theta(1+\cos^2\theta)\Big\}}
{r^2-2Mr+a^2}drd\phi ,
\label{LE}
\end{align}
where 
the coordinates are 
$-\infty < t < +\infty$, 
$2M \leq r < +\infty$, 
$0 \leq \theta \leq \pi$, 
$0 \leq \phi \leq 2\pi$ .
We denote 
\begin{align}
\beta^2=1-8\pi\eta^2, 
\label{beta2}
\end{align}
where $\eta$ is the scale of a gauge-symmetry breaking.

The rotating global monopole by Eq.(\ref{LE}) 
is a rotating generalization of the global monopole black hole
in Ref. \cite{GMP}. 
Here, $M$ denotes the global monopole core mass. 
The parameter $a$ is the total angular momentum of the global monopole,
which gives rise to the Lense-Thirring effect in general relativity, 
and the parameter $\beta$ is  
called the global monopole parameter 
of the spacetime 
where $\beta$ satisfies $0 < \beta \leq 1 $.

\subsection{Generalized optical metric}
By following Ref. \cite{OIA2017}, 
we define the generalized optical metric $\gamma_{ij}$ 
($i, j = 1, 2, 3$)
by a relation as 
\begin{align}
dt=& \sqrt{\gamma_{ij} dx^i dx^j} +\beta_i dx^i , 
\label{opt} 
\end{align}
which is directly obtained by solving the null condition ($ds^2 = 0$) 
for $dt$. 
Note that $\gamma_{ij}$ is not the induced metric 
in the Arnowitt-Deser-Misner (ADM) formalism. 
We define a three-dimensional space ${}^{(3)}M$ 
by the generalized optical metric $
\gamma_{ij}dx^idx^j$.

For the rotating global monopole by Eq. (\ref{LE}), 
we find the components of the generalized optical metric 
as 
\begin{align}
&\gamma_{ij}dx^idx^j \notag\\
=&\frac{\Big(a^2 \cos ^2\theta+r^2\Big)}
{\Big[a^2+r (r-2 M)\Big]^2 \Big[a^2 \cos ^2\theta+r (r-2 M)\Big]} \notag\\
&\times
\Big[a^4 \Big(\beta ^2-1\Big) \sin ^6\theta+a^2\{a^2+r(r-2M)\}\cos^2\theta
+a^2r^2\Big(\beta ^2-1\Big) \sin ^2\theta+ \Big(a^2-2 M r+r^2\Big)r^2\Big]dr^2 
\notag\\
&+\frac{\beta ^2 \left(a^2 \cos ^2\theta+r^2\right)^2}
{a^2 \cos ^2\theta+r (r-2 M)}d\theta^2
+\frac{2a \Big(1-\beta ^2\Big) 
\Big[r^2 \sin ^2\theta-a^2 \cos ^2\theta \Big(\cos ^2\theta+1\Big)\Big]}
{\Big[a^2+r (r-2 M)\Big] \Big(1-\frac{2 M r}{a^2 \cos ^2\theta+r^2}\Big)}drd\phi
\notag\\
&+\frac{\sin ^2\theta \Big(a^2 \cos (2 \theta )+a^2+2 r^2\Big)^2 
\Big[a^2 \Big(\beta ^2-1\Big) \cos (2 \theta )+a^2 \Big(\beta ^2+1\Big)
+2 \beta ^2 r (r-2 M)\Big]}{8 \Big[r (r-2 M)+a^2 \cos ^2\theta\Big]^2}d\phi^2 . 
\label{gamma}
\end{align}
We obtain the components of $\beta_i$ as 
\begin{align}
\beta_idx^i=&-\frac{2 a M r \sin ^2\theta }{a^2 \cos ^2\theta +r (r-2 M)}d\phi .
\end{align}

In the rest of the paper, we focus on the light rays 
in the equatorial plane, 
namely $\theta = \pi/2$. 
Note that the generalized optical metric $\gamma_{ij}$ doesn't mean 
an asymptotically flat space, because
there is the deficit angle of spacetime (if $\beta \neq 1$).

\section{Deflection angle of light by a rotating global monopole} 
\subsection{Deflection angle of light in asymptotically flat spacetimes}
Let us begin this section with briefly summarizing 
the generalized optical metric method that enables us to 
calculate the deflection angle of light 
for non-asymptotic receiver (denoted as $R$) 
and source (denoted as $S$) \cite{OIA2017}. 

We define the deflection angle of light as \cite{OIA2017}
\begin{equation}
\alpha \equiv \Psi_R - \Psi_S + \phi_{RS} . 
\label{alpha-axial}
\end{equation} 
Here, $\Psi_R$ and $\Psi_S$ are angles between the light ray tangent 
and the radial direction from the lens object, 
defined in a covariant manner using the generalized optical metric, 
at the receiver location and the source, respectively. 
On the other hand, 
$\phi_{RS}$ is the coordinate angle between the receiver and source, 
where the coordinate angle is associated with the rotational Killing vector 
in the spacetime. 
If the space under study is Euclidean, this $\alpha$ becomes 
the deflection angle of the curve. 
This is consistent with the thin lens approximation 
in the standard theory of gravitational lensing. 

By using the Gauss-Bonnet theorem as \cite{GBMath}
\begin{align}
\iint_{{}^{\infty}_{R}\square^{\infty}_{S}} K dS 
+ \oint_{\partial T} \kappa_g d\ell 
+ \sum_{i=1}^n \Theta_i = 2\pi .
\label{GBT}
\end{align}
Eq. (\ref{alpha-axial}) can be recast into \cite{OIA2017}
\begin{align}
\alpha 
=-\iint_{{}^{\infty}_{R}\square^{\infty}_{S}} K dS 
+ \int_{S}^{R} \kappa_g d\ell , 
\label{GB-axial}
\end{align} 
where 
$K$ is defined as the Gaussian curvature 
at some point on the two-dimensional surface, 
$dS$ denotes the infinitesimal surface element 
defined with $\gamma_{ij}$, 
${}^{\infty}_{R}\square^{\infty}_{S}$ denotes 
a quadrilateral embedded in a curved space with $\gamma_{ij}$, 
$\kappa_g$ denotes the geodesic curvature of the light ray in this space 
and $d\ell$ is an arc length defined with the generalized optical metric 
(See Fig. 2 in Ref. \cite{OIA2017}). 
It is shown by Asada and Kasai that 
this $d\ell$ for the light ray is an affine parameter \cite{AK}. 

\subsection{Deflection angle of light in spacetimes with
deficit angle}
When we consider the deflection angle of light in a spacetime 
with the deficit angle, we 
follow References \cite{Intro6,Intro7,OIA2017} 
to use 
the definition of deflection angle of light as
\begin{align}
\alpha \equiv \Psi_R - \Psi_S + \phi_{RS} . 
\label{modifyda}
\end{align}
In the rest of the present paper, 
we show that the deficit angle contribution 
to the deflection angle of light can be included.

Note that  the surface integral and path integral terms appear 
in the right hand side of Eq. (\ref{GB-axial}) if $\beta_i =0$ 
(See \cite{Intro6}).
However, in the rotating global monopole, 
Eq.(\ref{GB-axial}) is modified by the deficit angle.  
Eq. (\ref{GB-axial}) is calculated as 
\begin{align}
&\iint_{{}^{\infty}_{R}\Box^{\infty}_{S}} K dS 
+\int^{R}_{r_{\infty}} \tilde{\kappa_g} d\ell 
+\int^{r_{\infty}}_{S} \dot{\kappa_g} d\ell 
-\int_{C_r} \bar{\kappa_g} d\ell +\int_{C_{\infty}}\kappa_g d\ell 
+\Psi_R+(\pi-\Psi_S)+\pi=2\pi , 
\end{align}
which is rewritten as 
\begin{align}
&\iint_{{}^{\infty}_{R}\Box^{\infty}_{S}} K dS 
+\int^{R}_{r_{\infty}} \tilde{\kappa_g} d\ell 
+\int^{r_{\infty}}_{S} \dot{\kappa_g} d\ell 
-\int_{C_r} \bar{\kappa_g} d\ell +\beta \phi_{RS}
+\Psi_R-\Psi_S=0, 
\label{GB}
\end{align} 
where $\tilde{\kappa_g}$ is a geodesic curvature along 
the radial line from  the infinity to the receiver, 
$\dot{\kappa_g}$ is a geodesic curvature along the radial line from the source 
to the infinity, 
$\bar{\kappa_g}$ is a geodesic curvature along the light ray from the source 
to the receiver and
$\kappa_g$ is a geodesic curvature along the path $C_{\infty}$.
The path $C_r$ is a light ray from the receiver to the source 
in generalized optical metric, 
$C_{\infty}$ is a circular arcsegment of radius $R>>r_{R}, r_{S}$ and 
we use $d\ell =\sqrt{1+\frac{4M}{r}}dr=\{1+\mathcal{O}(M/r)\}dr$ 
along the radial line. 
We shall explain in more detail this calculation in Sec.III-D-3 .
Therefore, the deflection angle of light by the rotating global monopole is  
rewritten as 
\begin{align}
\alpha = -\iint_{{}^{\infty}_{R}\Box^{\infty}_{S}} K dS 
+\int^{r_{\infty}}_{R} \tilde{\kappa_g} d\ell 
-\int^{r_{\infty}}_{S} \dot{\kappa_g} d\ell 
+\int_{C_r}\bar{\kappa_g} d\ell 
+(1-\beta) \phi_{RS} , \label{GB-new}
\end{align}
where we use Eqs. (\ref{modifyda}) and (\ref{GB}). 
The deflection angle is also a coordinate-invariant 
in the spacetimes with deficit angle,
because $\Psi_R$ and $\Psi_S$ are obtained by the inner product at a receiver 
and a source respectively.

We have two ways in order to calculate the deflection angle of light.
We shall make detailed calculations of the right-hand side of 
Eq. (\ref{GB-new}) and the right-hand side of 
Eq. (\ref{modifyda})  below.

\subsection{Gaussian curvature}
For the equatorial case of a rotating global monopole, 
the Gaussian curvature in the weak field approximation is calculated as 
\begin{align}
K=&\frac{R_{r\phi r\phi}}{\det\gamma^{(2)}_{ij}}
\notag\\
=&\frac{1}{\sqrt{\det\gamma^{(2)}_{ij}}}\Big[\frac{\partial}{\partial\phi}
\Big(\frac{\sqrt{\det\gamma^{(2)}_{ij}}}{\gamma^{(2)}_{rr}}\Gamma^{\phi}_{~rr}\Big)
-\frac{\partial}{\partial r}
\Big(\frac{\sqrt{\det\gamma^{(2)}_{ij}}}{\gamma^{(2)}_{rr}}\Gamma^{\phi}_{~r\phi}\Big)\Big] 
\notag\\
=&\Bigg[-\frac{2}{r^3}-\frac{6}{r^5\beta^2}a^2\Bigg]M
+\frac{3}{r^4}M^2 \notag\\
&+\mathcal{O}(M^3/r^5),
\label{K}
\end{align}
where 
$\gamma^{(2)}_{ij}$ denotes the two-dimensional generalized optical metric 
in the equatorial plane $\theta = \pi/2$.  
Here, $a$ and $M$ are dimensional quantities that 
can be used as book-keeping symbols in iterative calculations 
under the weak field approximation. 
As for the first line of Eq. (\ref{K}), please see e.g. 
the page 263 in Reference \cite{LL}.  
We note that the first term in the second line of Eq. (\ref{K}) 
does not contribute because $\Gamma^{\phi}_{rr} = 0$. 
It is not surprising that this Gaussian curvature does not agree with 
Eq. (26) in 
Jusufi, Werner, Banerjee and \"Ovg\"un 
\cite{JO}, 
because their Gaussian curvature describes 
another surface that is associated with the Randers-Finsler metric 
different from our optical metric, though the same four-dimensional spacetime
is  considered by two groups.

In order to perform the surface integral of the Gaussian curvature 
in Eq. (\ref{GB-axial}), 
we have to know the boundary shape of the integration domain. 
In other words, we need to describe the light ray as a function of $r(\phi)$. 
For the later convenience, we introduce the inverse of $r$ as 
$u \equiv r^{-1}$. 
The orbit equation in this case becomes 
\begin{align}
&\bigg[1+2Mu\bigg]\bigg(\frac{d u}{d \phi}\bigg)^2
-\bigg[\frac{2 a \left(1-\beta ^2\right) 
\left(b^2u^2-\beta ^2\right)}{b^2}
-\frac{4 aMu \left(\beta ^2-1\right) 
\left(b^2u^2-2 \beta^2\right)}{b^2}\bigg]
\frac{d u}{d \phi} \notag\\
&+\bigg[\left(\beta^2u^2-\frac{\beta^4}{b^2}\right)
+\Big\{-\frac{2 \beta^4u}{b^2}+\frac{4 a \beta^4 u}{b^3}\Big\}M\bigg]
+\mathcal{O}(a^2u^2,M^2u^2)=0 , \label{OE}
\end{align}
where $b$ is the impact parameter of the photon. 
See e.g. Reference \cite{OIA2017} on how to obtain the photon 
orbit equation in the axisymmetric and stationary spacetime. 
The orbit equation
is iteratively solved as 
\begin{align}
u(\phi)=&\frac{\beta}{b}\sin\left\{\beta \phi+\phi_0(1-\beta)\right\}
+\left[\beta^2+\beta^2\cos^2\left\{\beta \phi+\phi_0(1-\beta)\right\}\right]
\frac{M}{b^2} \notag\\
&+\frac{\beta(\beta^2-1)
\sin[2\left\{\beta \phi+\phi_0(1-\beta)\right\}]}{2b^2}a \notag\\
&+\frac{\beta^2[-4+(-1+\beta^2)\cos\left\{\beta \phi+\phi_0(1-\beta)\right\}
+(-1+\beta^2)\cos\left\{3\beta \phi+3\phi_0(1-\beta)\right\}]}{2b^3}aM
\notag\\
&+\mathcal{O}(M^2/b^3).
\label{u}
\end{align}

The area element of the equatorial plane $dS$ is
\begin{align}
dS=\sqrt{\det\gamma^{(2)}_{ij}}drd\phi 
= \sqrt{\beta^2 r^2 + \mathcal{O}(Mr)}drd\phi
=\{\beta r + \mathcal{O}(M)\} dr d\phi .
\end{align}

By using Eq. (\ref{u}) as the iterative solution for the photon orbit, 
the surface integral of the Gaussian curvature in Eq. (\ref{GB-axial}) 
is calculated as 
\begin{align}
-\iint_{{}^{\infty}_{R}\square^{\infty}_{S}} K dS 
=&\int_{\infty}^{r(\phi)} dr
\int_{\phi_S}^{\phi_R} d\phi 
\left(-\frac{2M}{r^3} \right) r\beta
+\mathcal{O}(M^3/b^3,a^2M^3/b^5,a^4M^2/b^6)
\notag\\
=& \int_{0}^{u(\phi)}du\int^{\phi_R}_{\phi_S} d\phi (2M\beta) 
+\mathcal{O}(M^3/b^3,a^2M^3/b^5,a^4M^2/b^6)\notag\\
=& 2M\beta \int^{\phi_R}_{\phi_S} d\phi 
\left( \frac{\beta}{b}\sin\left\{\beta \phi+\phi_0(1-\beta)\right\}
+\frac{\beta(\beta^2-1)
\sin[2\left\{\beta \phi+\phi_0(1-\beta)\right\}]}{2b^2}a 
\right) \notag\\
=&\frac{2M\beta}{b}\left[\sqrt{1-\frac{b^2{u_S}^2}{\beta^2}}
+\sqrt{1-\frac{b^2{u_R}^2}{\beta^2}}\right]
+\frac{aM\left(1-\beta^2\right)}{\beta}\left[{u_R}^2-{u_S}^2\right] \notag\\
&+\mathcal{O}(M^3/b^3),
\label{intK}
\end{align}
where  
$u_R$ and $u_S$ are the inverse of $r_R$ and $r_S$, respectively 
and 
we used  
\begin{align}
\sin\{\beta\phi_S+\phi_0(1-\beta)\}=\frac{bu_S}{\beta}
+\frac{(1-\beta^2)\sqrt{1-\frac{b^2{u_S}^2}{\beta^2}}}{\beta}u_Sa
-\frac{\beta(2-\frac{b^2{u_S}^2}{\beta^2})}{b}M 
+\mathcal{O}(aM/b^2)
\end{align} 
and  
\begin{align}
\sin\{\beta\phi_R+\phi_0(1-\beta)\}=\frac{bu_R}{\beta}
-\frac{(1-\beta^2)\sqrt{1-\frac{b^2{u_R}^2}{\beta^2}}}{\beta}u_Ra
-\frac{\beta(2-\frac{b^2{u_R}^2}{\beta^2})}{b}M+\mathcal{O}(aM/b^2)
\end{align} 
by Eq. (\ref{u}) in the last line.

\subsection{Geodesic curvature}
\subsubsection{Light ray in optical metric}
The geodesic curvature plays an important role
in our calculations of the light deflection, though 
it is not usually described in standard textbooks on 
general relativity. 
Hence, we follow Reference \cite{OIA2017} to 
briefly explain the geodesic curvature here. 
The geodesic curvature can be defined in the vector form as 
(e.g. \cite{Math})
\begin{align}
\kappa_g \equiv \vec{T}^{\prime} \cdot \left(\vec{T} \times \vec{N}\right) , 
\label{kappag-vector}
\end{align}
where we assume a parameterized curve with a parameter $\ell$, 
$\vec{T}$ is the unit tangent vector for the curve 
by reparameterizing the curve using its arc length 
$\ell$, 
$\vec{T}^{\prime}$ is its derivative with respect to 
the arc length, 
and $\vec{N}$ is the unit normal vector for the surface. 
Eq. (\ref{kappag-vector}) can be rewritten in the tensor form as 
\begin{align}
\kappa_g = \epsilon_{ijk} N^i a^j e^k , 
\label{kappag-tensor}
\end{align}
where $\vec{T}$ and $\vec{T}^{\prime}$ are denoted by 
$e^k$ and $a^j$, respectively. 
Here, the Levi-Civita tensor 
$\epsilon_{ijk}$ is defined by 
$\epsilon_{ijk} \equiv \sqrt{\gamma}\varepsilon_{ijk}$, 
where 
$\gamma \equiv \det{(\gamma_{ij})}$, 
and $\varepsilon_{ijk}$ is the Levi-Civita symbol 
($\varepsilon_{123} = 1$). 
In the present paper,  we use $\gamma_{ij}$ in the above definitions 
but not $g_{ij}$. 
Note that $a^i \neq 0$ in the three-dimensional optical metric 
by nonvanishing $g_{0i}$ \cite{OIA2017}, 
even though the light signal follows a geodesic 
in the four-dimensional spacetime. 
On the other hand, we emphasize that 
$a^i = 0$ and thus $\kappa_g = 0$ 
for the geodesics in the optical metric, 
because $\beta_i = 0$. 

As shown first in Reference \cite{OIA2017}, 
Eq. (\ref{kappag-tensor}) is rewritten in a convenient form as 
\begin{align}
\kappa_g = - \epsilon^{ijk} N_i \beta_{j|k} , 
\label{kappag-tensor2} 
\end{align}
where we use $\gamma_{ij}e^ie^j = 1$. 

Let us denote the unit normal vector to the equatorial plane 
as $N_p$. 
Therefore, it satisfies 
$N_p \propto \nabla_p \theta = \delta^{\theta}_p$, 
where $\nabla_p$ is the covariant derivative associated with
$\gamma_{ij}$. 
Hence, $N_p$ is written in a form as 
$N_p = N_{\theta} \delta^{\theta}_p$. 
By noting that $N_p$ is a unit vector ($N_pN_q \gamma^{pq} = 1$), 
we obtain $N_{\theta} = \pm 1/\sqrt{\gamma^{\theta\theta}}$. 
Therefore, $N_p$ can be expressed as 
\begin{align}
N_p = \frac{1}{\sqrt{\gamma^{\theta\theta}}} \delta_p^{\theta} , 
\label{N}
\end{align}
where we choose the upward direction without loss of generality. 

For the equatorial case, one can show 
\begin{align}
\epsilon^{\theta p q} \beta_{q|p} 
&=-\frac{1}{\sqrt{\gamma}}\beta_{\phi,r} , 
\label{rot-beta}
\end{align}
where the comma denotes the partial derivative, 
we use $\epsilon^{\theta r \phi} = - 1/\sqrt{\gamma}$ 
and 
we note $\beta_{r,\phi} = 0$ owing to the axisymmetry. 
By using Eqs. (\ref{N}) and (\ref{rot-beta}),
the geodesic curvature of the light ray with the generalized optical metric 
becomes 
\cite{OIA2017}
\begin{align}
\kappa_g=-\sqrt{\frac{1}{\gamma\gamma^{\theta\theta}}}\beta_{\phi,r} . 
\label{kappa_equ}
\end{align}
For the global monopole case, this is obtained as 
\begin{align}
\bar{\kappa_g}=-\frac{2}{\beta r^3}aM-\frac{ 2 }{ \beta r^4 }aM^2
+\mathcal{O}(aM^3/r^5).
\end{align}

We examine the contribution from the geodesic curvature. 
This contribution is the path integral along the light ray 
(from the source to the receiver),  
which is computed as 
\begin{align}
\int_{C_r}\bar{\kappa}_gd\ell
=&-\int^{R}_{S}\frac{2}{\beta r^3}aM d\ell +\mathcal{O}(a^3M/b^4) 
\notag\\
=&-\int^{\phi_R}_{\phi_S}\frac{2}{\beta r^3}aM 
\frac{b}{\cos^2\{\beta\vartheta+\phi_0(1-\beta)\}}
d\vartheta +\mathcal{O}(aM^2/b^4) \notag\\
=&-\frac{2}{\beta}aM\int^{\phi_R}_{\phi_S} 
\left(\frac{\beta\cos\{\beta\vartheta+\phi_0(1-\beta)\}}{b}\right)^3
\frac{b}{\cos^2\{\beta\vartheta+\phi_0(1-\beta)\}}d\vartheta 
+\mathcal{O}(aM^2/b^4) \notag\\
=&-\frac{2\beta^2}{b^2}aM\int^{\phi_R}_{\phi_S} 
\cos\{\beta\vartheta+\phi_0(1-\beta)\} d\vartheta 
+\mathcal{O}(aM^2/b^4) \notag\\
=&-\frac{2aM\beta}{ b^2} 
[\sin\{\beta\phi_R+\phi_0(1-\beta)\}-\sin\{\beta\phi_S+\phi_0(1-\beta)\}]
+\mathcal{O}(aM^2/b^4) \notag\\
=&-\frac{2aM\beta}{ b^2} 
\left[
\sqrt{1-\frac{b^2{u_R}^2}{\beta^2}}+\sqrt{1-\frac{b^2{u_S}^2}{\beta^2}}
\right] +\mathcal{O}(aM^2/b^4),
\label{intkappa}
\end{align}
where we use $d\ell=
\frac{\beta^2r^2}{b} d\vartheta + \mathcal{O}(b^2/r^2, M)$, 
$
u 
=\frac{\beta\cos\{\beta\vartheta+\phi_0(1-\beta)\}}{b}+\mathcal{O}(a/b,M/b)$.
In the last line, we used 
$\sin\{\beta\phi_R+\phi_0(1-\beta)\}=\sqrt{1-\frac{b^2{u_R}^2}{\beta^2}}
+ \mathcal{O}(a u_R, M u_R)$ 
and $\sin\{\beta\phi_S+\phi_0(1-\beta)\}=-\sqrt{1-\frac{b^2{u_S}^2}{\beta^2}}
+ \mathcal{O}(a u_S, M u_S)$ by Eq. (\ref{u}). 
The sign of the right-hand side of Eq. (\ref{intkappa}) changes, 
if the photon orbit is retrograde. 

\subsubsection{Radial lines in the generalized optical metric}
The unit tangent vector along a radius line in ${}^{(3)}M$ is $R^i=(R^r,0,0)$. 
On the equatorial plane, from
\begin{align}
\gamma_{ij}R^iR^j=&\gamma_{rr}(R^r)^2=1, \notag\\
\end{align}
we obtain
\begin{align}
R^r=&\frac{1}{\sqrt{\gamma_{rr}}}. \label{radius}
\end{align}
The acceleration vector $a^i$ along this line is
\begin{align}
a^i=& {R^i}_{|j}R^j. \notag\\
\end{align}
Its explicit form is
\begin{align}
a^i=&\Bigg(\frac12\Big(\frac{\partial }{\partial r}\frac{1}{\gamma_{rr}}\Big)
+\frac{\gamma^{rr}}{2\gamma_{rr}}\frac{\partial \gamma_{rr}}{\partial r}
+\frac{\gamma^{r\phi}}{\gamma_{rr}}\frac{\partial \gamma_{r\phi}}{\partial r},
0,
\gamma^{\phi\phi}\frac{\partial \gamma_{r\phi}}{\partial r}
+\frac{\gamma^{r\phi}}{2}\frac{\partial \gamma_{rr}}{\partial r} \label{st_acce}
\Bigg).
\end{align}
Here, the vector $a^i$ $(i= r, \theta, \phi )$ becomes
\begin{align}
a^r= &\frac{2(\beta^2-1)^2}{\beta^2 r^4}a^2M +\mathcal{O}(a^4/r^5), \notag\\
a^{\theta}=&0, \notag\\
a^{\phi}=&\frac{2(\beta^2-1)}{\beta^2r^4}aM +\mathcal{O}(a^3/r^5,aM^2/r^5).
\end{align}
This means that $a^i$ is zero vector in Kerr or Schwarzschild cases 
($\beta=1$).

From Eq.(\ref{N}), we obtain 
\begin{align}
&N^i=\left(0,\frac{1}{\sqrt{\gamma_{\theta\theta}}},0\right). \label{normal}
\end{align}

By using Eqs.(\ref{kappag-tensor}), (\ref{radius}), (\ref{st_acce}) and 
(\ref{normal}), 
an explicit form of $\kappa_g$ is obtained as
\begin{align}
\kappa_g=&\epsilon_{ijk}N^ia^jR^k \notag\\
=&\sqrt{\frac{\gamma}{\gamma_{rr}\gamma_{\theta\theta}}}
\Bigg(\gamma^{\phi\phi}\frac{\partial \gamma_{r\phi}}{\partial r}
+\frac{\gamma^{r\phi}}{2}\frac{\partial \gamma_{rr}}{\partial r}\Bigg).
\label{radial-kappa}
\end{align}

Moreover, by substituting functions of metric $\gamma_{ij}$ 
into Eq.(\ref{radial-kappa}) , we obtain $\kappa_g$ as 
\begin{align}
\kappa_g=&-\sqrt{(r-2 M)^2 \Big\{a^2+r (r-2 M)\Big\} 
\Big\{a^2 \Big(\beta ^2-1\Big) (2 M r+1)+r^2\Big\}} \notag\\
&\times\Bigg[a \bigg(\beta ^2-1\bigg) r^2 \bigg\{a^4 \big(\beta ^2-1\big) 
\bigg(-8 M^2 r+M \big(3 r^2-5\big)+2 r\bigg)
+a^2 r \Big\{12 \big(\beta ^2-1\big) M^3 r \notag\\
&-8 \big(\beta ^2-1\big) M^2 \big(r^2-1\big)
+M r \big\{-6 \beta ^2+\big(\beta ^2-1\big) r^2+3\big\}+\beta ^2 r^2\big\}
+2 M r^3 (2 M-r)\Big\}\Bigg] \notag\\
&/\Bigg[(r-2 M)^2 \Big\{a^2+r (r-2 M)\Big\}^2 
\Big\{a^2 \big(\beta ^2-1\big) (2 M r+1)+r^2\Big\} \notag\\
&\times\Bigg\{r^2 \bigg(a^6 \big(\beta ^2-1\big) (2 M r+1)
+a^4 r \Big(-4 \big(\beta ^4-1\big) M^2 r
+2 \big(\beta ^4-1\big) M \big(r^2-1\big)+\beta ^4 r\Big) \notag\\
&+2 a^2 \beta ^2 r^2 (2 M-r) \Big(2 \big(\beta ^2-1\big) M^2 r
-\big(\beta ^2-1\big) M \big(r^2-1\big)-r\Big)
+\beta ^2 r^4 (r-2 M)^2\bigg)\bigg\}^{1/2} \Bigg].
\end{align}
This is approximated as
\begin{align}
\kappa_g=\frac{2 \left(\beta ^2-1\right)}{\beta r^3}aM
-\frac{ \beta\left(\beta ^2-1\right) }{r^4}a^3
+\frac{10(\beta^2-1)}{\beta r^4}aM^2
+\mathcal{O}(a^3M/r^5,aM^3/r^5),
\end{align}
where this $\kappa_g$  vanishes in Kerr or Schwarzschild spacetime $(\beta=1)$,
since the acceleration vector $a^i$ becomes $0$.

Let us integrate the leading term of $\kappa_g$ from the source to the infinity
\begin{align}
\int^{r_{\infty}}_{S}\frac{2 \left(\beta ^2-1\right)}{\beta r^3}aMd\ell=&
\int^{\infty}_{r_S}\frac{2 \left(\beta ^2-1\right)}{\beta r^3}aMdr \notag\\
=&\frac{ \left(\beta ^2-1\right)aM}{\beta}\Big[\frac{1}{r^2}\Big]^{r_S}_{\infty}
 \notag\\
=&-\frac{ \left(1-\beta^2\right)aM}{\beta {r_S}^2} 
+\mathcal{O}(aM^2/{r_{S}}^3). \label{SR1}
\end{align}
Similarly, the integral of $\kappa_g$ from the receiver to the infinity 
is computed as
\begin{align}
\int^{r_{\infty}}_{R}\frac{2 \left(\beta ^2-1\right)}{\beta r^3}aMd\ell
=&-\frac{ \left(1-\beta ^2\right)aM}{\beta {r_R}^2} 
+\mathcal{O}(aM^2/{r_{R}}^3), 
\label{SR2}
\end{align}
where we use $d\ell=\sqrt{1+\frac{4M}{r}}dr=\{1+\mathcal{O}(M/r)\}dr$ .

\subsubsection{Geodesic curvature of circular arcsegment in optical metric}

The orbital equation as Eq.(\ref{OE}) can be solved for $\frac{du}{d\phi}$ as
\begin{align}
\frac{du}{d\phi} =&\frac{1}{F_{\pm}(u)}, \notag\\
\end{align}
where we denote
\begin{align}
F_{+}(u)=&\frac{1}{\beta\sqrt{{u_0}^2-u^2}}-\left(1-\frac{1}{\beta^2}\right)a
+\frac{{u_0}^3- u^3 }{\beta({u_0}^2-u^2)^{3/2}}M \notag\\
&-2uaM\left(1-\frac{1}{\beta^2}
+\frac{{u_0}^3(u_0-u)}{u\beta^2 ({u_0}^2-u^2)^{3/2}}\right) \notag\\
&+\mathcal{O}(a^2u,M^2u,aM^2u^2,M^3u^2),\\
F_{-}(u)=&-\frac{1}{\beta\sqrt{{u_0}^2-u^2}}-\left(1-\frac{1}{\beta^2}\right)a
-\frac{{u_0}^3- u^3 }{\beta({u_0}^2-u^2)^{3/2}}M \notag\\
&-2uaM\left(1-\frac{1}{\beta^2}
-\frac{{u_0}^3(u_0-u)}{u\beta^2 ({u_0}^2-u^2)^{3/2}}\right) \notag\\
&+\mathcal{O}(a^2u,M^2u,aM^2u^2,M^3u^2).
\end{align}
For $\phi_0>\phi>\phi_S$, we use $F_+(u)$, while we use $F_-(u)$ for 
$\phi_R>\phi>\phi_0$.
Here, we use
\begin{align}
b=\frac{\beta}{u_0}+\beta M-2u_0aM
+\mathcal{O}(a^2{u_0},M^2{u_0},aM^2{u_0}^2,M^3{u_0}^2),
\end{align}
where $u_0$ is the inverse of the 
distance of closest approach. 

At $r=r_{\infty}$
($r_{\infty}$ is an infinite constant radius of the circular arc segment), 
we obtain $d\ell^2={r_{\infty}}^2\beta^2d\phi^2$, 
the geodesic curvature $\kappa_g=\frac{1}{r_{\infty}}+O(M/{r_{\infty}}^2)$.
Let us integrate as
\begin{align}
\beta \phi_{RS}=&\int^R_S\kappa_gdl=\int^{R}_{S} \beta d\phi
=\beta\int^{\phi_R}_{\phi_S}d\phi =\beta \int^{u_0}_{u_S} F_{+}(u)du 
+\beta\int^{u_R}_{u_0} F_{-}(u)du .
\end{align}
\begin{align}
\int F_{\pm}(u)du=&
\int\Big\{\pm\frac{1}{\beta\sqrt{{u_0}^2-u^2}}
-\left(1-\frac{1}{\beta^2}\right)a
\pm\frac{{u_0}^3- u^3 }{\beta({u_0}^2-u^2)^{3/2}}M \notag\\
&-2uaM\left(1-\frac{1}{\beta^2}
\pm\frac{{u_0}^3(u_0-u)}{u\beta^2 ({u_0}^2-u^2)^{3/2}}\right)\Big\}du \notag\\
=&\pm\frac{1}{\beta}\arcsin\left(\frac{u}{u_0}\right)
-\left(1-\frac{1}{\beta^2}\right)au
\mp\frac{(2u_0+u)\sqrt{{u_0}^2-u^2}}{\beta({u_0}+u)}M \notag\\
&+\left\{\left(-1+\frac{1}{\beta^2}\right)u^2
\pm\frac{2{u_0}^2\sqrt{{u_0}^2-u^2}}{\beta^2(u_0+u)}\right\}aM \notag\\
&+\mathcal{O}(M^2/{u_0}^2). 
\end{align}
\begin{align}
\phi_{RS}=&\frac{\pi}{\beta} -\frac{1}{\beta}
\left\{\arcsin\left(\frac{bu_S}{\beta}\right)
+\arcsin\left(\frac{bu_R}{\beta}\right)\right\}
-\left(1-\frac{1}{\beta^2}\right)(u_R-u_S)a \notag\\
&+\left\{
\frac{(2-\frac{b^2{u_R}^2}{\beta^2})}{b\sqrt{1-\frac{b^2{u_R}^2}{\beta^2}}}
+\frac{(2-\frac{b^2{u_S}^2}{\beta^2})}{b\sqrt{1-\frac{b^2{u_S}^2}{\beta^2}}}
\right\}M 
\notag\\
&+\left\{-\left(1-\frac{1}{\beta^2}\right)({u_R}^2-{u_S}^2)
-\frac{2}{b^2\sqrt{1-\frac{b^2{u_R}^2}{\beta^2}}}
-\frac{2}{b^2\sqrt{1-\frac{b^2{u_S}^2}{\beta^2}}}
\right\}aM +\mathcal{O}(M^2/b^2),
\label{PHIRS}
\end{align}
where we use 
$u_0 =\frac{\beta}{b} + \frac{\beta^2 M}{b^2} - \frac{2\beta^2aM}{b^3}$.
This $\phi_{RS}$ becomes that for the Kerr case, only if one takes the limit 
$\beta \rightarrow 1$.

\subsection{Jump angles}
In the previous section, the unit tangent vector along the radius line in 
${}^{(3)}M$ is  obtained as
\begin{align}
R^i=&(\frac{1}{\sqrt{\gamma_{rr}}},0,0), 
\end{align}
the unit tangential vector along the spatial curve is also obtained as 
\begin{align}
e^i =&\xi\Big(\frac{dr}{d\phi},0,1\Big) , \notag\\
\end{align}
where
\begin{align}
\xi_{+R}=&\frac{b}{{r_R}^2\beta^2} -\frac{2b}{{r_R}^3\beta^2}M
+\frac{(-1+\beta^2)\sqrt{1-\frac{b^2}{{r_R}^2\beta^2}}}{{r_R}^2\beta^2}a 
\notag\\
&+\frac{2{r_R}^2\beta^2(1-\frac{b^2}{{r_R}^2\beta^2})
+b^2(-1+\beta^2)\sqrt{1-\frac{b^2}{{r_R}^2\beta^2}}}
{{r_R}^5\beta^4(1-\frac{b^2}{{r_R}^2\beta^2})}aM +\mathcal{O}(M^2/{r_R}^3),
\notag\\
\xi_{-R}=& -\frac{b}{{r_R}^2\beta^2} +\frac{2b}{{r_R}^3\beta^2}M
-\frac{(-1+\beta^2)\sqrt{1-\frac{b^2}{{r_R}^2\beta^2}}}{{r_R}^2\beta^2}a 
\notag\\ 
&-\frac{2{r_R}^2\beta^2(1-\frac{b^2}{{r_R}^2\beta^2})
+b^2(-1+\beta^2)\sqrt{1-\frac{b^2}{{r_R}^2\beta^2}}}
{{r_R}^5\beta^4(1-\frac{b^2}{{r_R}^2\beta^2})}aM +\mathcal{O}(M^2/{r_R}^3),
\notag\\
\xi_{+S}=& \frac{b}{{r_S}^2\beta^2} -\frac{2b}{{r_S}^3\beta^2}M
-\frac{(-1+\beta^2)\sqrt{1-\frac{b^2}{{r_S}^2\beta^2}}}{{r_S}^2\beta^2}a 
\notag\\
&+\frac{2{r_S}^2\beta^2(1-\frac{b^2}{{r_S}^2\beta^2})
-b^2(-1+\beta^2)\sqrt{1-\frac{b^2}{{r_S}^2\beta^2}}}
{{r_S}^5\beta^4(1-\frac{b^2}{{r_S}^2\beta^2})}aM +\mathcal{O}(M^2/{r_S}^3)
,\notag\\
\xi_{-S}=& -\frac{b}{{r_S}^2\beta^2} +\frac{2b}{{r_S}^3\beta^2}M
+\frac{(-1+\beta^2)\sqrt{1-\frac{b^2}{{r_S}^2\beta^2}}}{{r_S}^2\beta^2}a 
\notag\\
&-\frac{2{r_S}^2\beta^2(1-\frac{b^2}{{r_S}^2\beta^2})
-b^2(-1+\beta^2)\sqrt{1-\frac{b^2}{{r_S}^2\beta^2}}}
{{r_S}^5\beta^4(1-\frac{b^2}{{r_S}^2\beta^2})}aM +\mathcal{O}(M^2/{r_S}^3) . 
\notag
\end{align}
Here, $\xi_{+}$ means that $e^i$ is the tangent vector of the 
prograde photon orbit and  $\xi_{-}$ means that $e^i$ is 
the tangent vector of the retrograde photon 
orbit. 
In addition, the subscripts S and R for $\xi_{\pm}$
mean from the source to the closest approach 
and from the receiver to the closest approach, respectively. 
Therefore, we can define the angle measured from the outgoing radial direction 
by
\begin{align}
\cos\Psi_R \equiv& \gamma_{ij}e^iR^j \notag\\
=&\gamma_{rr}e^rR^r+\gamma_{\phi r}e^{\phi}R^r \notag\\
=&\sqrt{\gamma_{rr}}\xi_{+R}\left.\frac{dr}{d\phi}\right|_+
+\frac{\gamma_{\phi r}}{\sqrt{\gamma_{rr}}}\xi_{+R} \notag\\
=&\sqrt{1-\frac{b^2}{{r_R}^2\beta^2}} 
+\frac{b^2M}{{r_R}^3\beta^2\sqrt{1-\frac{b^2}{{r_R}^2\beta^2}}}
+\frac{b(1-\beta^2)}{{r_R}^2\beta^2}a \notag\\
&-\frac{2b}{{r_R}^3\beta^2\sqrt{1-\frac{b^2}{{r_R}^2\beta^2}}}aM 
+\mathcal{O}(M^2/{r_R}^2),\label{cosR}\\
-\cos(\pi-\Psi_S) \equiv& \gamma_{ij}e^iR^j \notag\\
=&\gamma_{rr}e^rR^r+\gamma_{\phi r}e^{\phi}R^r \notag\\
=&\sqrt{\gamma_{rr}}\xi_{+S}\left.\frac{dr}{d\phi}\right|_{-}
+\frac{\gamma_{\phi r}}{\sqrt{\gamma_{rr}}}\xi_{+S} \notag\\
=&-\sqrt{1-\frac{b^2}{{r_S}^2\beta^2}} 
-\frac{b^2M}{{r_S}^3\beta^2\sqrt{1-\frac{b^2}{{r_S}^2\beta^2}}}
+\frac{b(1-\beta^2)}{{r_S}^2\beta^2}a \notag\\
&+\frac{2b}{{r_S}^3\beta^2\sqrt{1-\frac{b^2}{{r_S}^2\beta^2}}}aM 
+\mathcal{O}(M^2/{r_S}^2), \label{cosS}
\end{align}
where Eq. (\ref{cosR}) is at the receiver position 
and Eq.(\ref{cosS}) is at the source.
Therefore, $\Psi_R$ and $\Psi_S$ are obtained as
\begin{align}
\Psi_R=&\arcsin\left(\frac{b}{{r_R}\beta}\right)
-\frac{bM}{{r_R}^2\beta\sqrt{1-\frac{b^2}{{r_R}^2\beta^2}}}
+\frac{(\beta^2-1)a}{{r_R}\beta} \notag\\
&+\frac{2+(\beta^2-1)\sqrt{1-\frac{b^2}{{r_R}^2\beta^2}}}
{{r_R}^2\beta\sqrt{1-\frac{b^2}{{r_R}^2\beta^2}}}aM 
+\mathcal{O}(M^2/{r_R}^2), \label{PSIR}\\
\pi-\Psi_S=&\arcsin\left(\frac{b}{{r_S}\beta}\right)
-\frac{bM}{{r_S}^2\beta\sqrt{1-\frac{b^2}{{r_S}^2\beta^2}}}
-\frac{(\beta^2-1)a}{{r_S}\beta} \notag\\
&+\frac{2-(\beta^2-1)\sqrt{1-\frac{b^2}{{r_S}^2\beta^2}}}
{{r_S}^2\beta\sqrt{1-\frac{b^2}{{r_S}^2\beta^2}}}aM 
+\mathcal{O}(M^2/{r_S}^2). \label{PSIS}
\end{align}

\subsection{Deflection angle}
By bringing together 
Eqs. (\ref{intK}), (\ref{intkappa}), (\ref{SR1}), (\ref{SR2}), 
(\ref{PSIR}) and (\ref{PSIS}), 
the deflection angle of light for the prograde case is obtained as 
\begin{align}
\alpha_{\mbox{prog}}
=& \left(\frac{1}{\beta}-1\right)\pi 
-\left(\frac{1}{\beta}-1\right)
\left\{\arcsin\left(\frac{b^2{u_R}^2}{\beta^2}\right)
+\arcsin\left(\frac{b^2{u_S}^2}{\beta^2}\right)\right\} \notag\\
&+\frac{(\beta-1)^2(\beta+1)}{\beta^2}(u_R-u_S)a 
+\left\{
\frac{2\beta-(1+\frac{1}{\beta})b^2{u_R}^2}{b\beta\sqrt{1-\frac{b^2{u_R}^2}
{\beta^2}}}
+\frac{2\beta-(1+\frac{1}{\beta})b^2{u_S}^2}{b\beta\sqrt{1-\frac{b^2{u_S}^2}
{\beta^2}}}\right\}M \notag\\
&-\left\{
\frac{2(\beta-b^2{u_R}^2)}{b^2\beta\sqrt{1-\frac{b^2{u_R}^2}{\beta^2}}}
+\frac{2(\beta-b^2{u_S}^2)}{b^2\beta\sqrt{1-\frac{b^2{u_S}^2}{\beta^2}}}
-\frac{(\beta-1)^2(\beta+1)}{\beta^2}({u_R}^2-{u_S}^2)
\right\}aM \notag\\
&+\mathcal{O}(M^2/b^2). 
\label{alpha-prog}
\end{align}

The deflection angle for the retrograde case is 
\begin{align}
\alpha_{\mbox{retro}}=&
\left(\frac{1}{\beta}-1\right)\pi 
-\left(\frac{1}{\beta}-1\right)
\left\{\arcsin\left(\frac{b^2{u_R}^2}{\beta^2}\right)
+\arcsin\left(\frac{b^2{u_S}^2}{\beta^2}\right)\right\} \notag\\
&-\frac{(\beta-1)^2(\beta+1)}{\beta^2}(u_R-u_S)a 
+\left\{
\frac{2\beta-(1+\frac{1}{\beta})b^2{u_R}^2}{b\beta\sqrt{1-\frac{b^2{u_R}^2}
{\beta^2}}}
+\frac{2\beta-(1+\frac{1}{\beta})b^2{u_S}^2}{b\beta\sqrt{1-\frac{b^2{u_S}^2}
{\beta^2}}}\right\}M \notag\\
&+\left\{
\frac{2(\beta-b^2{u_R}^2)}{b^2\beta\sqrt{1-\frac{b^2{u_R}^2}{\beta^2}}}
+\frac{2(\beta-b^2{u_S}^2)}{b^2\beta\sqrt{1-\frac{b^2{u_S}^2}{\beta^2}}}
-\frac{(\beta-1)^2(\beta+1)}{\beta^2}({u_R}^2-{u_S}^2)
\right\}aM \notag\\
&+\mathcal{O}(M^2/b^2).  
\label{alpha-retro}
\end{align}
If $\beta =1$, Eqs. (\ref{alpha-prog}) and (\ref{alpha-retro}) 
agree with the known result for the weak field approximation 
of the Kerr spacetime in Reference \cite{OIA2017}. 
For both cases, the source and receiver may be located 
at finite distance from the monopole. 
As a matter of course, these results are also obtained by substituting 
Eqs.(\ref{PHIRS}), (\ref{PSIR}) and (\ref{PSIS}) to Eq.(\ref{alpha-axial}).
Eqs. (\ref{alpha-prog}) and (\ref{alpha-retro}) show that 
the light deflection is affected by deficit angle.

One can see that, in the limit as $r_R \to \infty$ and $r_S \to \infty$, 
Eqs. (\ref{alpha-prog}) and (\ref{alpha-retro}) become 
\begin{align}
\alpha_{\mbox{prog}} \rightarrow &
\left(\frac{1}{\beta}-1\right)\pi 
+\frac{4M}{b}-\frac{4aM}{b^2} 
+\mathcal{O}\left(\frac{M^2}{b^2}\right) \notag\\
=& \left(\frac{1}{\beta}-1\right)\pi 
+\frac{4M}{b_K \beta}-\frac{4aM}{(b_K \beta)^2} 
+\mathcal{O}\left(\frac{M^2}{{b_K}^2}\right) \notag\\
=& \left(\frac{1}{\beta}-1\right)\pi 
+\frac{4M}{b_K}+\frac{16\pi \eta^2 M}{b_K}
-\frac{4aM}{{b_K}^2} -\frac{32\pi \eta^2 aM}{{b_K}^2}
+\mathcal{O}\left(\frac{M^2}{{b_K}^2}\right),
\label{alpha-infinity-prog}\\
\alpha_{\mbox{retro}} \rightarrow &
 \left(\frac{1}{\beta}-1\right)\pi 
+\frac{4M}{b_K}+\frac{16\pi \eta^2 M}{b_K}
+\frac{4aM}{{b_K}^2} +\frac{32\pi \eta^2 aM}{{b_K}^2}
+\mathcal{O}\left(\frac{M^2}{{b_K}^2}\right), 
\label{alpha-infinity-retro}
\end{align}
where $b_K$ is a constant of integration in 
Jusufi, Werner, Banerjee and \"Ovg\"un 
\cite{JO}, 
we used $b_K=b/\beta$ and $\beta^2=1-8\pi \eta^2$. 
These equations coincide with Eq.(53) in  
\cite{JO}, 
in which they are restricted within the asymptotic source and receiver 
($r_R \to \infty$ and $r_S \to \infty$). 
Note that 
Reference \cite{JO} 
obtained $\pm\frac{128\pi \eta^2 aM}{5{b_K}^2}$ 
by their method, while a method of 
the direct integration of the null geodesic gives 
$\pm\frac{32\pi \eta^2 aM}{{b_K}^2}$:  
The former expression agrees with the latter one 
but with a different numerical coefficient.
Please see Appendix A of Reference \cite{JO}, 
especially Eq. (53) and the last paragraph of the appendix. 
According to their comments in the last paragraph, 
their approximation would need to 
be modified to recover a correct expression as 
$\pm\frac{32\pi \eta^2 aM}{{b_K}^2}$. 
Our result as Eqs. 
(\ref{alpha-infinity-prog}) and (\ref{alpha-infinity-retro}) 
is indeed in agreement with the latter expression. 
In this sense, our present approach is better than 
the method 
in Ref. 
\cite{JO}.

\section{Possible astronomical applications}
In this section, we discuss possible astronomical applications.
The above calculations discuss the deflection angle of light.
In particular, we do not assume 
that the receiver and the source are located at the infinity. 
The finite-distance correction to the deflection angle of light, 
denoted as $\delta\alpha$, is the difference between 
the asymptotic deflection angle $\alpha_{\infty}$ and the deflection angle 
for the finite distance case. 
It is expressed as 
\begin{align}
\delta\alpha \equiv \alpha_{\infty}-\alpha .
\end{align}
The finite-distance correction to the deflection angle of light 
is roughly estimated as 
\begin{align}
\delta\alpha \sim & 
\left(\frac{1}{\beta}-1\right)
\left\{\arcsin\left(\frac{b^2{u_R}^2}{\beta^2}\right)
+\arcsin\left(\frac{b^2{u_S}^2}{\beta^2}\right)\right\} \notag\\
&-\frac{(\beta-1)^2(\beta+1)}{\beta^2}(u_R-u_S)a \notag\\
&+\left\{
\frac{2\beta(\sqrt{1-\frac{b^2{u_R}^2}{\beta^2}}-1)
+(1+\frac{1}{\beta})b^2{u_R}^2}{b\beta\sqrt{1-\frac{b^2{u_R}^2}
{\beta^2}}}
+\frac{2\beta(\sqrt{1-\frac{b^2{u_R}^2}{\beta^2}}-1)
+(1+\frac{1}{\beta})b^2{u_S}^2}{b\beta\sqrt{1-\frac{b^2{u_S}^2}
{\beta^2}}}\right\}M \notag\\
&+\mathcal{O}(aM/b^2,M^2/b^2) \notag\\
\sim &\left(\frac{1}{\beta}-1\right)
\left\{\arcsin\left(\frac{b^2{u_R}^2}{\beta^2}\right)
+\arcsin\left(\frac{b^2{u_S}^2}{\beta^2}\right)\right\} \notag\\
&-\frac{(\beta-1)^2(\beta+1)}{\beta^2}(u_R-u_S)a 
+\left\{\frac{b{u_R}^2}{\beta}
+\frac{b{u_S}^2}{\beta}\right\}M \notag\\
&+\mathcal{O}(aM/b^2,M^2/b^2). 
\label{finite-collc}
\end{align}
The counterpart for the weak-field and slow-rotation Kerr metric is 
\cite{OIA2017}
\begin{align}
\delta\alpha_{Kerr} \sim 
(b u_R^2 + b u_S^2) M + \mathcal{O}(aM/b^2, M^2/b^2) . 
\label{finite-Kerr}
\end{align}
From Eqs. (\ref{finite-collc}) and (\ref{finite-Kerr}), 
the finite correction to the light deflection purely due to the angle deficit 
$\delta\alpha - \delta\alpha_{Kerr}$ 
becomes 
\begin{align}
\delta\alpha - \delta\alpha_{Kerr} 
\sim & 
\left(\frac{1}{\beta}-1\right)
\left\{\arcsin\left(\frac{b^2{u_R}^2}{\beta^2}\right)
+\arcsin\left(\frac{b^2{u_S}^2}{\beta^2}\right)\right\} \notag\\
&-\frac{(\beta-1)^2(\beta+1)}{\beta^2}(u_R-u_S)a 
+ \left(\frac{1}{\beta}-1\right) 
b({u_R}^2 + {u_S}^2) M \notag\\
&+\mathcal{O}(aM/b^2,M^2/b^2). 
\label{finite-deficit}
\end{align}

\begin{figure}[h]
\includegraphics[width=12cm]{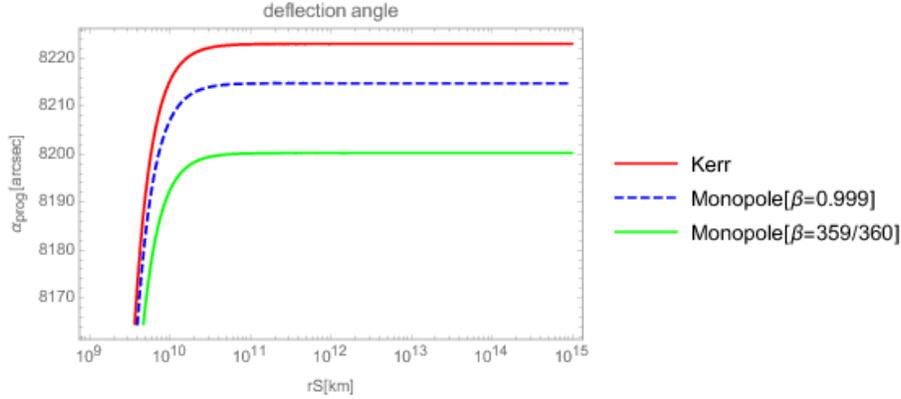}
\caption{$\alpha_{prog}$, where we assume the SgrA${}^{*}$. 
The vertical axis denotes the deflection angle of light with 
the finite-distance correction and 
the horizontal axis denotes the source distance $r_S$. 
The red solid curve, blue dash curve and green dot curve correspond to 
$\beta= 0$(Kerr spacetime), $\beta=0.999$ and $\beta=359/360$, respectively. 
The impact parameter is assumed to be $b=10^2 M_{Sgr}$. }
\label{fig-alpha-prograde}
\end{figure}

\begin{figure}[h]
\includegraphics[width=12cm]{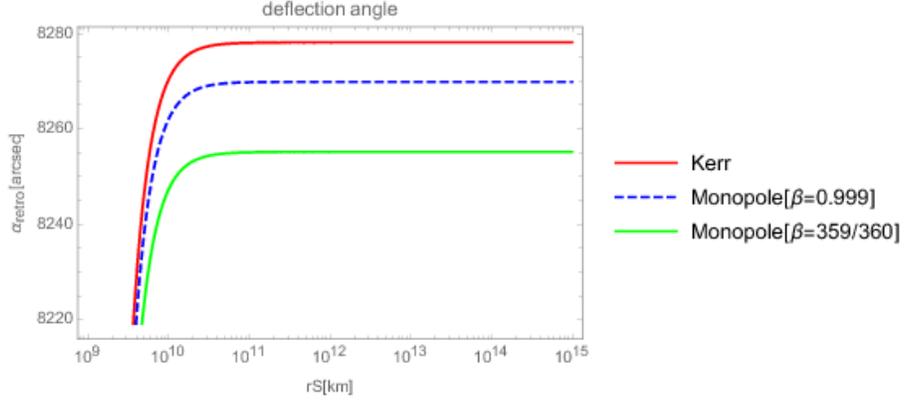}
\caption{$\alpha_{retro}$, where we assume the SgrA${}^{*}$. 
The vertical axis denotes the deflection angle of light with 
the finite-distance correction and 
the horizontal axis denotes the source distance $r_S$. 
The red solid curve, blue dash curve and green dot curve correspond to 
$\beta= 0$(Kerr spacetime), $\beta=0.999$ and $\beta=359/360$, respectively. 
The impact parameter is assumed to be $b=10^2 M_{Sgr}$.}
\label{fig-alpha-retrograde}
\end{figure}

For its simplicity,  
we consider the mass of the rotating global monopole equals to Sgr A${}^{*}$ 
($M_{Sgr} \simeq 4\times10^{6}M_{\odot}$, $M_{\odot}$ is Solar mass), 
the spin angular momentum of the rotating global monopole is 
$a=2/3 M_{Sgr}$ and the parameter $\beta=0,~ 0.999,~359/360$ .
We assume $r_R$ is the distance from Earth to SgrA${}^{*}$
($r_R \simeq 8 \times 10^3 [\mbox{pc}]$). 
We also assume $b \sim 100 M$ and $r_s \sim 0.1 \mbox{pc}$. 
As a rough order-of-magnitude estimate under these assumptions, 
three terms in Eq. (\ref{finite-deficit}) become 
\begin{align}
&\left(\frac{1}{\beta}-1\right)
\left\{\arcsin\left(\frac{b^2{u_R}^2}{\beta^2}\right)
+\arcsin\left(\frac{b^2{u_S}^2}{\beta^2}\right)\right\} 
\notag\\
&
\sim 
8 \times 10^{-3} \left( \frac{1-\beta}{10^{-3}} \right) 
\left( \frac{b}{100M_{Sgr}} \right)^2 
\left( \frac{0.1 \mbox{pc}}{r_S} \right)^2 
[\mbox{mas}] , 
\label{term-1}
\end{align}
\begin{align}
- \frac{(\beta-1)^2(\beta+1)}{\beta^2}(u_R-u_S)a 
\sim 
5 \times 10^{-4} 
\left( \frac{1-\beta}{10^{-3}} \right)^2 
\left( \frac{0.1 \mbox{pc}}{r_S} \right) 
\left( \frac{a}{2M/3} \right) 
[\mbox{mas}] , 
\label{term-2}
\end{align}
\begin{align}
\left(\frac{1}{\beta}-1\right) 
b({u_R}^2 + {u_S}^2) M 
\sim 
8 \times 10^{-5}
\left( \frac{1-\beta}{10^{-3}} \right) 
\left( \frac{b}{100M_{Sgr}} \right)
\left( \frac{0.1 \mbox{pc}}{r_S} \right)^2
\left( \frac{M}{4 \times 10^6 M_{\odot}} \right)
[\mbox{mas}] , 
\label{term-3}
\end{align}
respectively. 
The second and third terms are thus beyond reach of the present technology. 
On the other hand, the first term is much larger than the second and third ones 
and it may be probed by using the present technology, 
if $\beta$ is large enough. 
If present and near-future observations at the level of 
$\sim 1 \times 10^{-3}$ [mas]  
find no evidence of the first term, 
an upper bound on $1-\beta$ will be placed by Eq. (\ref{term-1}) as 
$1-\beta < \frac18 \times 10^{-3} \sim 1 \times 10^{-4}$. 
For the deficit angle $\delta = 8\pi^2\eta^2 = \pi (1-\beta^2)$, 
this bound is interpreted as 
$\delta \sim 2\pi(1-\beta) < 8 \times 10^{-4}$ [rad], 
where we use $1+\beta \sim 2$ for the small angle deficit. 
Figure \ref{fig-alpha-prograde} 
shows the gravitational deflection of light in the prograde orbit for 
SgrA${}^{*}$. 
Figure \ref{fig-alpha-retrograde} 
shows that for the retrograde orbit.

\afterpage{\clearpage}
\newpage

\section{Conclusion}
In the weak field approximation, 
we have discussed the deflection angle of light 
for an observer and source 
at finite distance from a rotating global monopole with deficit angle. 
We have shown that both of 
the Werner's method and the generalized optical metric method 
give the same deflection angle at the leading order of 
the weak field approximation, 
if the receiver and source are at the null infinity. 
Therefore, our result is a possible extension 
to asymptotically nonflat spacetimes.
We have also found corrections for the deflection angle 
due to the finite distance from the global monopole.  
We examined whether near-future observations of Sgr A${}^{*}$ 
can put an upper bound on the deficit angle 
for a rotating global monopole model. 
It is left for future to study higher order terms 
in the weak field approximation 
of a rotating global monopole and to examine also the strong deflection limit. 

\begin{acknowledgments}
We are grateful to Marcus Werner for the useful discussions. 
We wish to thank Kimet Jusufi for his comments for 
clarifying his paper Ref. \cite{JO}. 
We would like to thank 
Yuuiti Sendouda, Ryuichi Takahashi, Yuya Nakamura and 
Naoki Tsukamoto 
for the useful conversations. 
This work was supported 
in part by Japan Society for the Promotion of Science (JSPS) 
Grant-in-Aid for Scientific Research, 
No. 17K05431 (H.A.),  No. 18J14865 (T.O.), 
in part by Ministry of Education, Culture, Sports, Science, and Technology,  
No. 17H06359 (H.A.)
and 
in part by JSPS research fellowship for young researchers (T.O.).  
\end{acknowledgments}


\begin{thebibliography}{99}
\bibitem{Intro1}
 A. Einstein, Ann. Phys. (Berlin) {\bf 354}, 769 (1916).
\bibitem{Intro2}
 F. W. Dyson, A. S. Eddington, and C. Davidson, Phil. Trans.
R. Soc. A {\bf 220}, 291 (1920).
\bibitem{Intro3}
Y. Hagihara, Jpn. J Astron. Geophys. 8, 67 (1931).;
 S. Chandrasekhar, The Mathematical Theory of Black Holes
(Oxford University Press, New York, 1998).;
 C. W. Misner, K. S. Thorne, and J. A. Wheeler, Gravitation
(Freeman, New York, 1973).;
 C. Darwin, Proc. R. Soc. A 249, 180 (1959).;
 V. Bozza, Phys. Rev. D 66, 103001 (2002).;
 S. V. Iyer and A. O. Petters, Gen. Relativ. Gravit. 39, 1563
(2007).;
 V. Bozza and G. Scarpetta, Phys. Rev. D 76, 083008
(2007).;
 S. Frittelli, T. P. Kling, and E. T. Newman, Phys. Rev. D 61,
064021 (2000).;
 K. S. Virbhadra and G. F. R. Ellis, Phys. Rev. D 62, 084003
(2000).;
 K. S. Virbhadra, Phys. Rev. D 79, 083004 (2009).;
 K. S. Virbhadra, D. Narasimha, and S. M. Chitre, Astron.
Astrophys. 337, 1 (1998).;
 K. S. Virbhadra and G. F. R. Ellis, Phys. Rev. D 65, 103004
(2002).;
 K. S. Virbhadra and C. R. Keeton, Phys. Rev. D 77, 124014 (2008).;
 S. Zschocke, Classical Quantum Gravity 28, 125016 (2011).;
\bibitem{Intro4}
E. F. Eiroa, G. E. Romero, and D. F. Torres, Phys. Rev. D 66, 024010 (2002).;
V. Perlick, Phys. Rev. D 69, 064017 (2004).;
F. Abe, Astrophys. J. 725, 787 (2010).;
Y. Toki, T. Kitamura, H. Asada, and F. Abe, Astrophys. J. 740, 121 (2011).;
K. Nakajima and H. Asada, Phys. Rev. D 85, 107501 (2012).;
G. W. Gibbons and M. Vyska, Classical Quantum Gravity 29, 065016 (2012).;
J. P. DeAndrea and K. M. Alexander, Phys. Rev. D 89, 123012 (2014).;
T. Kitamura, K. Nakajima, and H. Asada, Phys. Rev. D 87, 027501 (2013).;
N. Tsukamoto and T. Harada, Phys. Rev. D 87, 024024 (2013).;
K. Izumi, C. Hagiwara, K. Nakajima, T. Kitamura, and H. Asada, 
Phys. Rev. D 88, 024049 (2013).;
T. Kitamura, K. Izumi, K. Nakajima, C. Hagiwara, and H. Asada, 
Phys. Rev. D 89, 084020 (2014).;
K. Nakajima, K. Izumi, and H. Asada, Phys. Rev. D 90,
084026 (2014).;
N. Tsukamoto, T. Kitamura, K. Nakajima, and H. Asada, 
Phys. Rev. D 90, 064043 (2014).;
\bibitem{Intro5}
G. W. Gibbons and M. C. Werner, Class. Quantum Grav. {\bf 25}, 235009 (2008).
\bibitem{Intro6}
A. Ishihara, Y. Suzuki, T. Ono, T. Kitamura and H. Asada, Phys. Rev. D {\bf 94},
 084015 (2016).
\bibitem{Intro7}
A. Ishihara, Y. Suzuki, T. Ono and H. Asada, Phys. Rev. D {\bf 95},
 044017 (2017).
\bibitem{Intro8}
M. Barriola and A. Vilenkin, Phys. Rev. Lett. {\bf 63}, 341
(1989).
\bibitem{Intro9}
B. Bertrand, Y. Brihaye and B. Hartmann, Class. Quan-
tum Grav. 20 4495 (2003); Y. Brihaye, B. Hartmann and
E. Radu, Phys. Rev. D 74, 025009 (2006).
\bibitem{Intro10}
A. Achucarro, B. Hartmann and J. Urrestilla, JHEP {\bf 0507}, 006 (2005).
\bibitem{Intro11}
H. Cheng and J. Man, Class. and Quantum Grav. {\bf 28},
015001 (2011).
\bibitem{Intro12}
L.-H. Liu and T. Prokopec, arXiv:1612.00861 [gr-qc]
\bibitem{RGMP}
R. M. Teixeira Filho and V. B. Bezerra, Phys. Rev D {\bf 64}, 084009 (2001).
\bibitem{GMP}
A. Vilenkin and E. P. S. Shellard, 
Cosmic Strings and Other Topological Defects 
(Cambridge University Press, Cambridge, 1994).
\bibitem{OIA2017}
T. Ono, A. Ishihara, and H. Asada, Phys. Rev. D {\bf 96}, 104037 (2017).
\bibitem{GBMath}
M. P. Do Carmo, {\it Differential Geometry of Curves and Surfaces}, 
pages 268-269, (Prentice-Hall,
New Jersey, 1976).
\bibitem{AK}
H. Asada, and M. Kasai, Prog. Theor. Phys. {\bf 104}, 95 (2000).
\bibitem{LL}
L. D. Landau, E. M. Lifschitz, The Classical Theory of Fields 
(Third English Edition), (Pergamon Press, Oxford, 1971).
\bibitem{JO}
K. Jusufi, M. C. Werner, A. Banerjee, A. \"Ovg\"un, 
Phys. Rev. D {\bf 95}, 104012 (2017)
\bibitem{Math}
A. C. Belton, Geometry of Curves and Surfaces, page 38 (2015);
www.maths.lancs.ac.uk/ belton/www/notes/geom notes.pdf; 
J. Oprea, Differential Geometry and Its Applications (2nd Edition), 
page 210, (Prentice Hall, New Jersey, 2003).
\end{thebibliography}
\end{document}